\documentclass[twocolumn,preprintnumbers,showpacs,amsmath,amssymb]{revtex4}

\usepackage{graphicx}
\usepackage{dcolumn}
\usepackage{bm}
\usepackage{epsfig}
\usepackage{amssymb}
\begin{document}

\preprint{}

\title{Comment on ``Fitting the annual modulation in DAMA with neutrons from muons and neutrinos"}

\author{P.S. Barbeau$^{a}$, J.I. 
Collar$^{b}$, Yu. Efremenko$^{c}$, and K. Scholberg$^{a,*}$.
}
\address{ 
$^{a}$Department of Physics, Duke University, Durham, NC 27708 USA\\
$^{b}$Department of Physics,
University of Chicago, Chicago, IL 60637 USA\\
$^{c}$University of Tennessee, Knoxville, TN 37996 USA\\
$^{*}${\frenchspacing Corresponding author. E-mail: schol@phy.duke.edu}\\
}
\begin{abstract}
We estimate rates of solar neutrino-induced neutrons in a DAMA/LIBRA-like detector setup, and find that the needed contribution to explain the annual modulation would require neutrino-induced neutron cross sections several orders of magnitude larger than current calculations indicate.  Although these cross sections have never been measured, it is likely that the solar-neutrino effect on DAMA/LIBRA is negligible.
\end{abstract}
\pacs{95.35.+d, 13.15.+g, 25.30.Pt, 26.65.+t}

\maketitle

In a recent Letter \cite{davis}, a new mechanism is proposed to explain the modulation effect 
apparent in DAMA/LIBRA dark matter detector data. Neutrons induced by $^{8}$B 
solar neutrinos (solar NINs) would add to those produced by muon interactions in the vicinity of the detectors, generating a modulation with a phase matching that observed by DAMA/LIBRA. The modulation amplitude required from this newly-considered process is 0.039 counts per kg of NaI[Tl] per day in the 2-6 keVee spectral region, for interactions affecting single DAMA/LIBRA crystals, which is comparable to what would be needed from muon-induced neutrons \cite{davis}. The origin of this modulated background is the eccentricity of the Earth's orbit, which changes the local $^{8}$B neutrino flux annually by $\pm 3.3$\%. This implies a large mean rate of $R_{\nu}\!\sim$11.5 NIN interactions per day in the 2-6 keVee singles spectrum from each individual 9.7 kg NaI[Tl] DAMA/LIBRA detector module. 

In this Comment we focus on the proposed solar NIN effect, which is the new idea presented in~\cite{davis}, although there are 
additional issues with the scenario; for example, the 
contribution from muon-induced 
neutrons has already been shown to fall short by three orders of magnitude in explaining the DAMA/LIBRA modulation \cite{damamuons} and is two orders of magnitude too small for a low-significance modulation in CoGeNT data \cite{cogentlong}. 
Leaving these aside, three overoptimistic approximations have been made in the estimation of the signal rate due to NINs in \cite{davis}. First, an effective volume of target material $V\sim$ 1000 m$^{3}$ is considered. This is much larger than the existing volume of high-NIN-cross-section material near
the detector;
only $\sim$ 0.98 m$^{3}$ of lead is present in the DAMA/LIBRA shield \cite{nimdama}. Second, the NIN mean free path is taken to be 2.6 m, justifying the large $V$ considered. While this is correct for energetic O(100) MeV neutrons generated by muon interactions, NINs should carry O(1) MeV energies characteristic of an evaporation spectrum \cite{spectrum}, resulting in a smaller range (few tens of cm in concrete), and modest $V$. Third, \cite{davis} does not consider  the small efficiency for transporting NINs from their originating site through shielding to the DAMA/LIBRA modules, and then generating single crystal interactions restricted to the 2-6 keVee spectral energy region. 

Assuming a standard $^8$B solar neutrino flux~\cite{bahcall}, neutrino oscillations consistent with  $\nu_e$ survival probability of $\sim$0.4, 
and using~\cite{snowglobes}, which employs cross sections from \cite{engel}, 
we calculate the rate of solar-neutrino-induced single-neutron emission in $^{208}$Pb
to be 0.85 per kilotonne of lead per day.
This rate is approximately consistent with the back-of-the-envelope calculation from~\cite{davis}.

We have performed two independent calculations of this NIN transport, one based on GEANT 4.10.0.1, the second using MCNP-PoliMi, both in agreement. The geometry of the detectors and shielding (copper, lead, cadmium, polyethylene) follows \cite{nimdama}. One meter of Gran Sasso cavern concrete and two of rock are included. For the results reported here, the geometry uses 10 cm of polyethylene moderator (up to 40 cm are present \cite{nimdama}), to allow a maximum of external NINs to reach the detectors. Quenching factors from \cite{quenching} are used to generate ionization energy (keVee) spectra from NIN interactions affecting only one of the sixteen external DAMA/LIBRA detectors.  These simulations indicate that the DAMA/LIBRA solar NIN modulation would extend to several tens of keVee, as opposed to the observed 2-6 keVee range. 

We find that only $\sim$0.046\% of the NINs generated in lead deposit energy in the 2-6 keVee range of a given external detector module, without involving interactions with next-neighbor crystals.  This results in $4.3\times 10^{-6}$ NINs per day creating relevant signals, which in turn implies that the NIN cross section in lead that would be required to produce the necessary $R_{\nu}$ is more than six orders of magnitude greater than calculated in~\cite{engel}.     We would require a cross section nearly as large for NINs in the copper surrounding the detector. 
Given the self-shielding, the presence of polyethylene moderator, and additional distance to the detector,
the solar NIN cross section in the simulated 114 m$^3$ of concrete
necessary to explain DAMA/LIBRA observations would have to be more than seven orders of magnitude larger than that for lead, for the thesis in \cite{davis} to hold. 
The contribution from 660 m$^{3}$ of rock is negligible, due to moderation in the innermost 1 m of concrete.

\begin{figure}
\includegraphics[width=8cm]{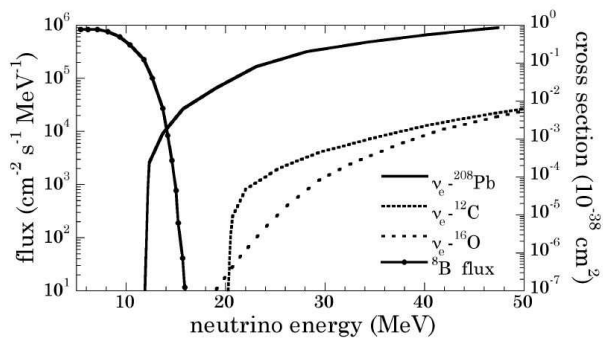}
\caption{Sample of charged-current neutrino cross-sections \protect\cite{snowglobes,engel}. Neutron-generating partial cross-sections involve higher thresholds, and have a magnitude smaller by a factor of a few. Neutral-current equivalents are smaller by an order of magnitude.  We considered only lead in our calculations;  the $^{8}$B solar neutrino flux is not energetic enough to produce NIN reactions in many of the lighter targets making up much of the shielding material. }
\end{figure}

The pertinent cross sections (Fig.\ 1) have never been measured.
We have recently embarked on an effort to measure them at the Spallation Neutron Source \cite{coherent1,coherent2}, although
we do not expect a discrepancy with respect to theoretical predictions \cite{engel,spectrum} sufficient to account for the DAMA/LIBRA modulation.


\end{document}